\documentclass[aps,pra,twocolumn,showpacs,preprintnumbers,amsmath,amssymb,floatfix,superscriptaddress]{revtex4-1}

\usepackage{graphics}
\usepackage{graphicx}
\usepackage{amsmath}
\usepackage{setspace}
\usepackage{braket}
\usepackage{epstopdf}
\usepackage{comment}
\usepackage{hyperref}
\usepackage{bm}
\usepackage{xfrac}
\usepackage{xcolor}

\usepackage{mathrsfs}

\hypersetup{%
   pdfpagemode=None, 
   pdfstartpage=1,
   pdfmenubar=true,
   pdftoolbar=true,
   colorlinks = true,
   linkcolor=blue,
   citecolor=blue,
   urlcolor=blue,
   bookmarksopen=false
 }

\begin{document}
\author{Miruna T. Cretu}
\address{Fritz-Haber-Institut der Max-Planck-Gesellschaft, Faradayweg 4-6, D-14195 Berlin, Germany}
\author{Marjan Mirahmadi}
\address{Fritz-Haber-Institut der Max-Planck-Gesellschaft, Faradayweg 4-6, D-14195 Berlin, Germany}
\author{Jes\'us P\'{e}rez-R\'{i}os}
\address{Fritz-Haber-Institut der Max-Planck-Gesellschaft, Faradayweg 4-6, D-14195 Berlin, Germany}
\address{Department of Physics, Stony Brook University, Stony Brook, New York 11794, USA}
\address{Institute for Advanced Computational Science, Stony Brook University, Stony Brook, NY 11794-3800, USA}

\title{Ion-atom-atom three-body recombination in cold hydrogen and deuterium plasmas}


\begin{abstract}

We present a detailed study about ion-atom-atom three-body recombination in hydrogen and deuterium plasmas based on classical trajectory calculations in hyperspherical coordinates. Our results, due to the predominant role of the long-range charged-induced dipole interaction, indicate that H$_2^+$ and D$_2^+$ are the main reaction products in the case of hydrogen and deuterium plasmas, respectively. Besides, we find a more steep energy-dependent reaction rate when the collision energy surpasses the dissociation energy of the molecular ion, thus entering a new dynamical regime dominated by short-range interactions.

\end{abstract}

\maketitle

\section{Introduction}

Three-body recombination is the chemical process in which three atoms collide to form a molecule as a product state. These reactions are relevant in many different areas of physics and chemistry. For instance, in ultracold gases they play a decisive role on the stability of a quantum gas\cite{Esry1999,Weiner1999,Bedaque2000,Suno2003,Weber2003,Schmidt2020,Greene2017}; in hybrid atom-ion traps they lead to new charged products~\cite{Perez-Rios2021,Hirzler2020,Perez-Rios2015,Krukow2016}; in chemical physics, they are necessary to understand the formation of van der Waals molecules~\cite{Brahms2008,Suno2009,Brahms2010,Brahms2011,Wang2011,Tariq2013,Quiros2017,Mirahmadi2021,Mirahmadi2021b}; in astrophysics they are responsible for the formation of H$_2$, one of the most important coolants regarding star formation~\cite{Flower2007,Forrey2013}. In the same vein, in plasmas, where ions, electrons, and neutral atoms live together, three-body recombination of electrons and protons, e + e + H$^+\rightarrow$ H + e, plays a vital influence as one of the main mechanisms of plasma recombination~\cite{Krasheninnikov1997,krasheninnikov2017,KUKUSHKIN2017984}. 

Among different three-body processes, ion-atom-atom three-body recombination is especially interesting since it involves two possible product states: (i) molecular ion formation, A + A + A$^+\rightarrow$ A$_2^+$ + A; (ii) molecular formation, A + A + A$^+\rightarrow$ A$_2$ + A$^+$\footnote{Please note that in general, ion-atom-atom three-body recombination is given by B + B + A$^+$.}. Recently, this reaction has been studied by applying a classical trajectory method in hyperspherical coordinates~\cite{Perez-Rios2014,Greene2017,Perez-Rios2020} finding a threshold law \cite{Perez-Rios2015} showing the dominance of molecular ion formation concerning neutral molecules\cite{Perez-Rios2018,Perez-Rios2021} at low collision energies. These have been experimentally confirmed in hybrid atom-ion traps experiments~\cite{Krukow2016,Mohammadi2021}. However, in the plasma physics front, despite the presence of ions and neutrals, the study of ion-atom-atom three-body recombination has received little theoretical and experimental attention barring the work of Kresti\'c et al~\cite{Krsti2003}.

This work presents a first-principles calculation of the ion-atom-atom three-body recombination rate in hydrogen and deuterium plasmas in a wide range of temperatures relevant to cold plasma physics. A classical trajectory method in hyperspherical coordinates is employed, based on the ideas pioneered by Smith\cite{Smith1962}. Monte Carlo (MC) sampling of trajectories is used to calculate the opacity function --the reaction probability as a function of the collision energy and impact parameter-- and the three-body recombination cross-section, which is conducive to calculating the recombination rate. The study led to the expected preponderance of molecular ion formation to the detriment of neutral molecules, although molecules are formed in the whole range of temperatures under consideration. These results are further discussed in the framework of tokamak physics, particularly its implications regarding the ultimate divertor detachment stage.

\section{Theoretical treatments and computational details}

Consider a system of three particles with masses $m_i$ and positions $\vec{r}_i$ ($i=1,2,3$) interacting via the potential $V( \vec{r_1},\Vec{r_2},\vec{r_3}) = U(r_{12})+U(r_{23})+U(r_{31})$, where $r_{ij}=|\vec{r}_j-\vec{r}_i|$. The Hamiltonian governing the particles' motion reads as

\begin{equation}
\label{eqn:hamiltonian}
    H = \frac{\vec{p}_1^2}{2m_1} + \frac{\vec{p}_2^2}{2m_2} + \frac{\vec{p}_3^2}{2m_3} + V(\vec{r}_1,\vec{r}_2,\vec{r}_3),
\end{equation}

\noindent where $\vec{p}_i$ are the momentum vectors of the $i$\textsuperscript{th} particle. To more conveniently solve Hamilton's equations and study the classical scattering of the three particles, the problem is treated in Jacobi coordinates:

\begin{eqnarray}
\label{eqn:jacobi}
    \vec{\rho}_1 &= &\vec{r}_2 - \vec{r}_1, \nonumber \\
    \vec{\rho}_2 &= &\vec{r}_3 - \frac{m_1\vec{r}_1 + m_2\vec{r}_2}{m_1 + m_2}, \nonumber \\
    \vec{\rho}_{CM} &=& \frac{m_1\vec{r}_1 + m_2\vec{r}_2 + m_3\vec{r}_3}{M},
\end{eqnarray}

\noindent where $M = m_1+m_2+m_3$ is the total mass of the system and $\vec{\rho}_{CM}$ is the three-body center-of-mass vector. As the total linear momentum is conserved, the degrees of freedom of the center of mass are neglected and the Hamiltonian from Eq.~(\ref{eqn:hamiltonian}) is expressed as

\begin{equation}
\label{eqn:hamiltonian_jacobi}
    H = \frac{\vec{P}_1^2}{2\mu_{12}} + \frac{\vec{P}_2^2}{2\mu_{3,12}} +  V(\vec{\rho}_1,\vec{\rho}_2).
\end{equation}

\noindent Here, $\mu_{12}=m_1m_2/(m_1+m_2)$, $\mu_{3,12}=m_3(m_1+m_2)/M$ and $\vec{P}_1$, $\vec{P}_2$ indicate the conjugated momenta of $\vec{\rho_1}$ and $\vec{\rho}_2$, respectively. $V(\vec{\rho}_1,\vec{\rho}_2)$ is the interaction potential expressed in terms of Jacobi coordinates. 

Solving Hamilton's equations of motion for the three-body problem as a function of Jacobi coordinates allows the characterization of the system in three-dimensional (3D) space. Alternatively, it is desirable to study the dynamics of the same system in a six-dimensional (6D) space, where it can be mapped into a single particle. This is made possible by the pioneering work of Smith\cite{Smith1962}, which allows the construction of 6D position and momenta vectors. Hyperspherical coordinates are employed for the representation of the 6D vectors and a definition of the impact parameter in 6D space follows. That is, the impact parameter $\vec{b}$ is defined as a component of the position vector lying in the hyperplane perpendicular to the initial momentum $\vec{P}_0$ and is obtained as a result of the Gram-Schmidt orthogonalization procedure.

Thus, an expression for the classical cross section, as a prerequisite for the recombination rate relation, is obtained after averaging out the hyperangular degrees of freedom~\cite{Greene2017,Perez-Rios2020}:

\begin{equation}
\label{eqn:cross_section}
\begin{split}
    \sigma_{rec}(E_c) = \int \mathcal{P}(E_c,\vec{b})b^4dbd\Omega_b \\
    = \frac{8\pi^2}{3} \int^{b_{max}(E_c)}_0 \mathcal{P}(E_c,b)b^4db,
\end{split}
\end{equation}

\noindent where $E_c$ is the collision energy and $d\Omega_b=\sin^3(\alpha_4^b)\sin^2(\alpha_3^b)\sin(\alpha_2^b)d\alpha_4^bd\alpha_3^bd\alpha_2^bd\alpha_1^b$ is the solid angle element associated with the vector $\vec{b}$. The function $\mathcal{P}(E_c,b)$ is the so-called opacity function and it defines the probability that a particular trajectory at collision energy $E_c$ and impact parameter $b$ leads to a recombination event. $b_{max}$ represents the largest impact parameter for which three-body recombination events occur, i.e. $\mathcal{P}(E_c,b)=0$ for $b>b_{max}$.
Finally, the energy-dependent three-body recombination rate is introduced as

\begin{equation}
\label{eqn:rate}
    k_3(E_c)=\sqrt{\frac{2E_c}{\mu}}\sigma_{rec}(E_c).
\end{equation}

Therefore, after starting with a given set of initial conditions $\vec{\rho}_0$, $R$, $\vec{b}$ and $\vec{P}_0$, the method relies on mapping the conditions in 3D space to solve Hamilton's equations in Jacobi coordinates. Following that, the resulting momenta and positions are transformed back into the 6D space where the classical three-body cross section is defined by Eq.~(\ref{eqn:cross_section}).

The integral in Eq.~(\ref{eqn:cross_section}) is evaluated by employing the Monte Carlo integration method~\cite{Perez-Rios2014}: the initial hyperangles determining the orientation of vectors $\vec{P}_0$ and $\vec{b}$ in 6D space are randomly sampled from their appropriate angular distribution functions in 6D and the opacity function $\mathcal{P}(E_c,\vec{b})$ is subsequently averaged over the hyperangles. This is achieved by running $n_t$ trajectories and counting the number of trajectories $n_r$ that lead to recombination events. Thus,

\begin{equation}
    \label{eqn:opacity}
    \mathcal{P}(E_c,b) \approx \frac{n_r(E_c,b)}{n_t(E_c,b)}\pm \frac{\sqrt{n_r(E_c,b)}}{n_t(E_c,b)}\sqrt{\frac{n_t(E_c,b)-n_r(E_c,b)}{n_t(E_c,b)}},
\end{equation}

\noindent where the latter term in Eq.~(\ref{eqn:opacity}) is the statistical error associated with the Monte Carlo method.

\begin{figure}[t]
 \includegraphics[width=1\linewidth]{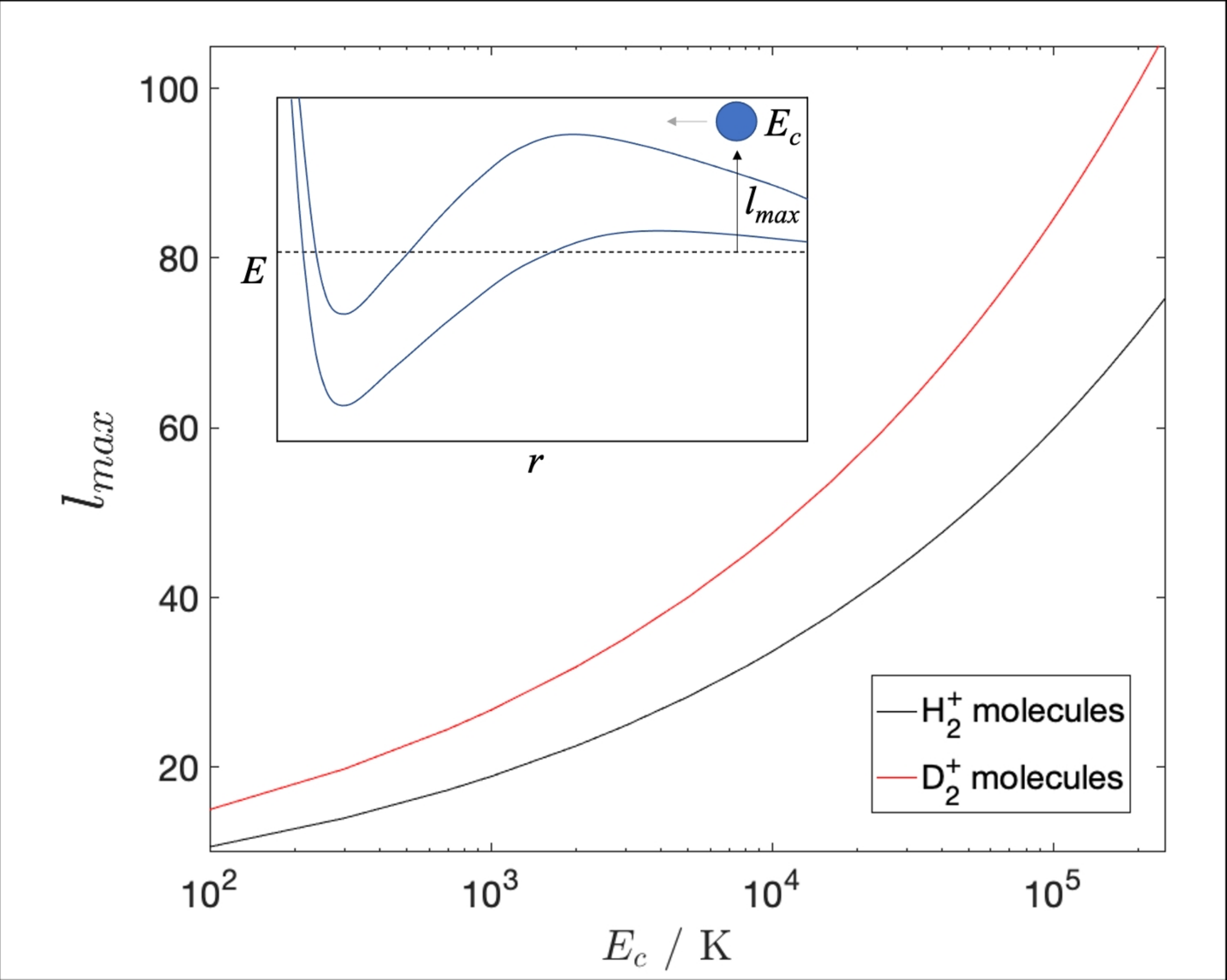} 
\caption{\label{fig:partial_waves} Maximum number of partial waves ($l_{max}$) vs. collision energy ($E_c$) as calculated for a Langevin impact parameter. The concept figure in the inset is illustrative of the maximum contributing partial wave to an ion-atom collision. }
\end{figure}


\begin{figure*}[t]
\begin{center}
 \includegraphics[width=0.93\linewidth]{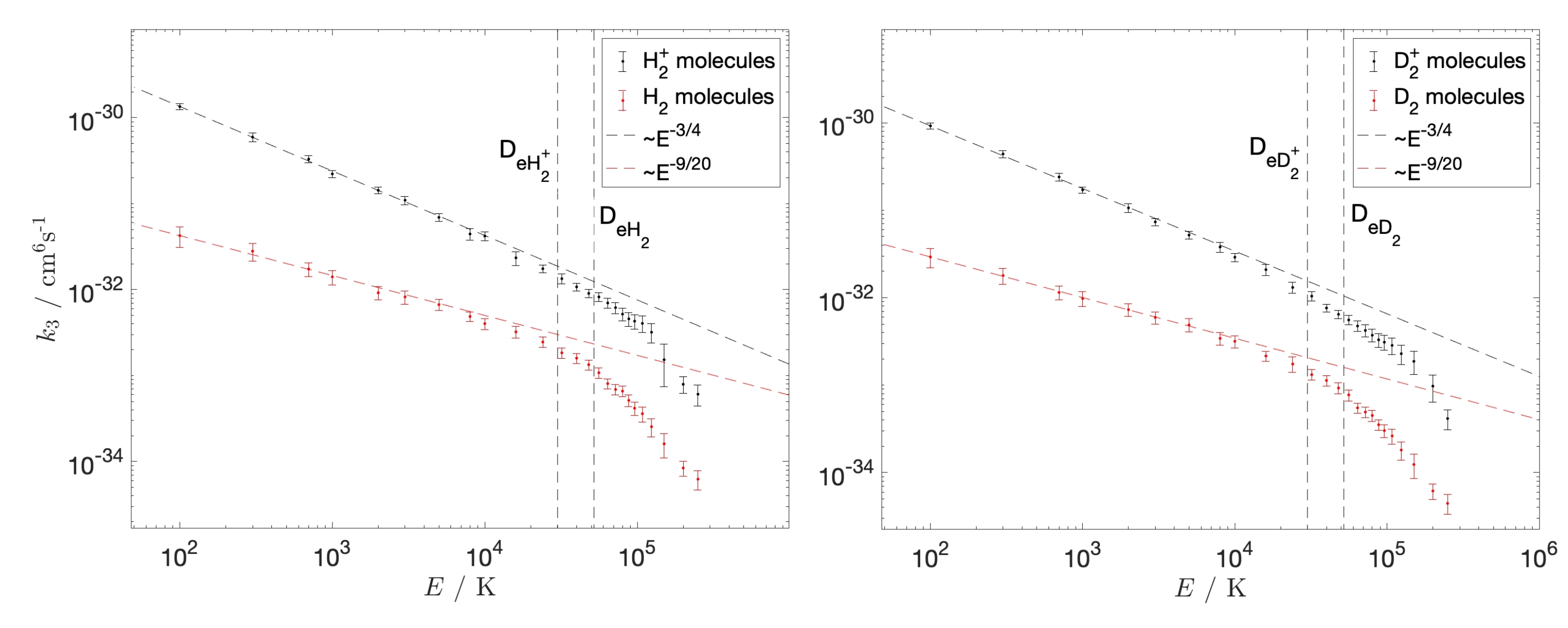}
\caption{\label{fig:k3_vs_en} Energy-dependent ion-atom-atom three-body recombination rates for the hydrogen and deuterium plasmas. The black symbols depict the rate of molecular ion formation, whereas the red symbols describe the formation of molecules. The error bars attached to each symbol account for one standard deviation error as customary in Monte Carlo simulations [see Eq.~(\ref{eqn:opacity})]. The vertical lines in the left panel represent the dissociation energies of H$_2^+$ and H$_2$, whereas, in the right panel, the vertical lines represent the same observable for D$_2^+$ and D$_2$.}
\end{center}
\end{figure*}

Hamilton's equations are solved using the ode113 solver in Matlab, with absolute and relative tolerances of $10^{-15}$ and $10^{-11}$, respectively. The conservation of the total energy and of the magnitude of the total angular momentum vector during collisions is ensured to at least four and six significant digits, respectively. Last but not least, the initial magnitude of the hyperradius ($\vec{\rho}_0$) is generated randomly from the interval $[R_0-50,R_0+50]$ $a_0$, where $R_0=40$ $a_0$ and $a_0$ is the Bohr radius. $R_0$ is chosen so that the three particles are initially found in an uniform rectilinear state of motion.

The number of partial waves contributing to the scattering observables can be used as a measure of the validity for the classical trajectory method. By knowledge of the long range of the interaction potential ($V_l$) we solve $E_c = V_l(R^*) $ for a given collision energy $E_c$, with $R^*$ being the distance where $V_l$ shows a maximum. Therefore, the largest $l$ value can be approximated, which determines the number of partial waves contributing to the scattering. For atom-ion collisions, the long range of the interaction potential reads as (in atomic units):

\begin{equation}
    V_l(R) = -\frac{\alpha}{2R^4} + \frac{l(l+1)}{2\mu R^2},
\end{equation}

\noindent where $l$ is the angular momentum quantum number associated to the relative motion, $\alpha$ is the polarisability of the atom and $\mu$ is the reduced mass of the colliding pair. Fig. \ref{fig:partial_waves} shows the magnitude of the maximum $l$ value ($l_{max}$) as a function of collision energy $E_c$, where we observe that at least 10 partial waves contribute to the observables at the lowest collision energy we consider later in this study ($E_c = 100$ K). As a result, quantum mechanical effects are potentially washed out for the range of collision energies explored in this work, and a classical approach is adequate.


\begin{figure*}[t]
\begin{center}
 \includegraphics[width=0.9\linewidth]{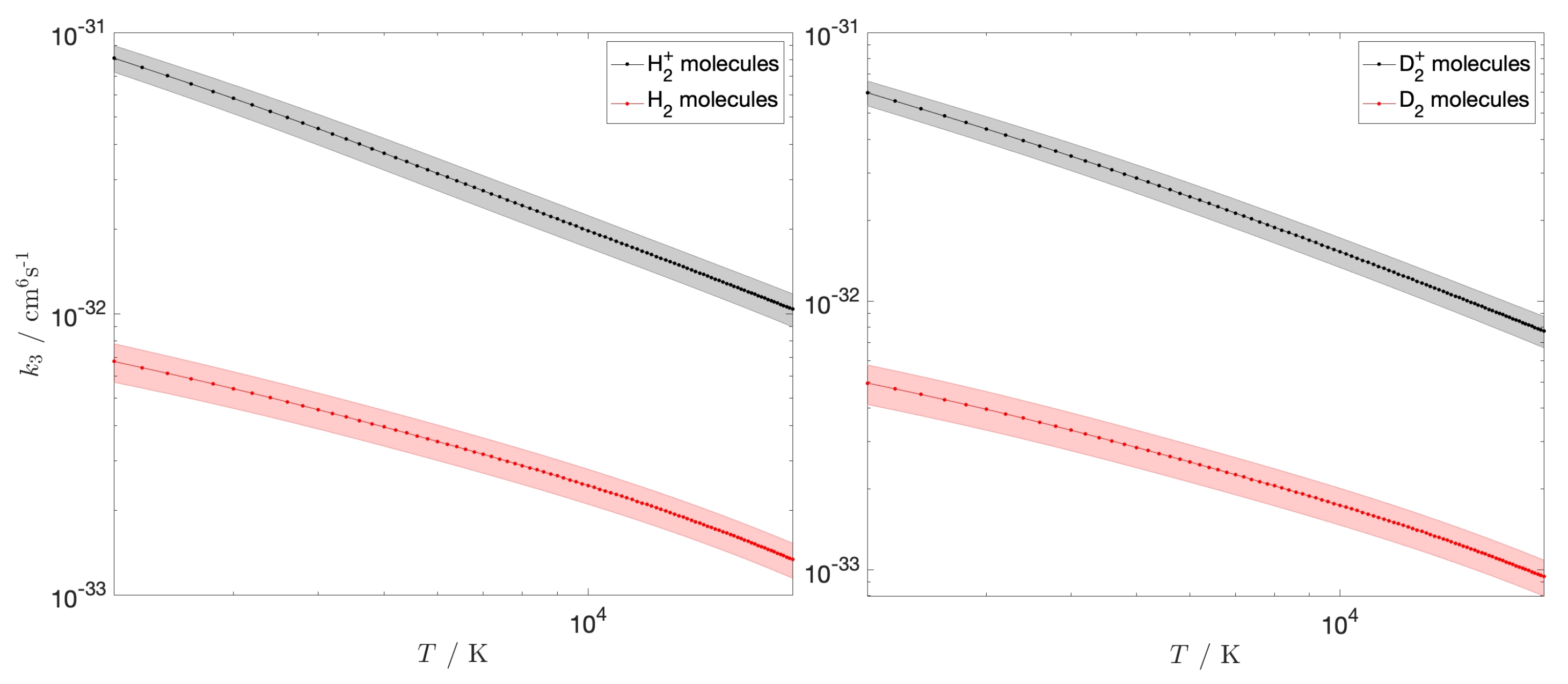} 
\caption{\label{fig:thavg}Thermally averaged ion-atom-atom three-body recombination rates for hydrogen (left panel) and deuterium plasmas (right panel). The black symbols depict the rate of molecular ion formation, whereas the red symbols describe the formation of molecules. The shades account for one standard deviation error as customary in Monte Carlo simulations [see Eq.~(\ref{eqn:opacity})].} 
\end{center}
\end{figure*}

\section{Results and discussion}

Three-body recombination processes involving hydrogen and deuterium species are considered simultaneously throughout this section.

The pairwise additive approximation is used to describe the overall potential of the three-body systems under study. A similar approach has been proven successful in cold chemistry~\cite{Krukow2016,Mohammadi2021,Hirzler2022}, ion mobility experiments~\cite{Perez-Rios2018} and van der Waals molecular formation~\cite{Mirahmadi2021}. Here, atom-atom interactions are modeled by a Lennard-Jones potential: $U(r) = C_{12}/r^{12} - C_6/r^6$, with $C_6=6.499$ a.u.\cite{Yan1996} and dissociation energy $D_e=36118.11$ cm\textsuperscript{-1} ($D_e=36748.38$ cm\textsuperscript{-1} for deuterium) \cite{Jinjun2009} that correlates with the X$^1\Sigma_g$ electronic state. Ion-atom interactions are described by the model potential $U(r) = C_8/r^8 - C_4/r^4$, that correlates with the X$^2\Sigma_g^+$ electronic state with dissociation energy $D_e=21379.35$ cm\textsuperscript{-1}($D_e=21711.64$ cm\textsuperscript{-1} for deuterium)\cite{Jinjun2009} and $C_4=\alpha/2$, where $\alpha$ denotes the static dipole polarizability of hydrogen ($\alpha=9/2$ a.u.\cite{polarisabilities2019}). 

The energy dependent ion-atom-atom three-body recombination rates $k_3(E_c)$ are depicted in Fig.~\ref{fig:k3_vs_en}, where a factor of $1/4$ has been included to account for the different spin states of the H$^+_3$ interaction~\cite{Krsti2003}. Notably, $k_3(E_c)$ is almost two orders of magnitude larger for the formation of the molecular ion compared to that of the molecule, which is expected given the stronger nature of interactions for the former. Two power-law behaviours are identified for each species formation, and the change in trend is established around the dissociation energy $D_e$ of the molecule or molecular ion products. The same behaviour was noted for the formation of van der Waals molecules X-RG (where RG is a rare gas atom) through three-body recombination\cite{Mirahmadi2021}. Here, Mirahmadi et al. showed that the change in trend could be attributed to an interplay between the contributions of the X-RG potential's short- and long-range tails to the molecule formation, depending on the energy of the system. Based on this, we distinguish low-energy and high-energy regimes in Fig. \ref{fig:k3_vs_en}, with a steeper trend characterizing the high-energy region. In that respect, the long-range tail of the potential plays an important part in the formation of molecules at low collision energies, and the opposite is true for the short-range tail.

A power law for the molecular ion recombination rate at low energies was inferred from Fig. \ref{fig:k3_vs_en}, yielding:

\begin{equation}
    k_3\sim E^{-3/4}.
\end{equation}

\noindent In fact, this low-energy power law behaviour for the formation of molecular ions through three-body recombination was devised in multiple other systems~\cite{Perez-Rios2020} and was also derived by Smirnov in 1967, who treated three-body recombination as a two-step chemical process~\cite{smirnov1967}.

In a different vein, the rate for molecule formation follows a less steep trend and is defined by the following power law

\begin{equation}
    k_3\sim E^{-9/20},
\end{equation}

\noindent
which, it is very close to the expected behavior for a three-body recombination process dominated by a potential $\propto r^{-5}$. However, any of the interatomic potentials included in our calculations show this long-range behavior.   

Furthermore, it is worth investigating the thermal averaged three-body recombination rate at temperatures $2000$ K $\leq T \leq 20000$ K. The thermal average is obtained via integrating the energy-dependent three-body recombination rate in Eq.~(\ref{eqn:rate}) over the appropriate Maxwell-Boltzmann distribution of collision energies:

\begin{equation}
    k_3(T) = \frac{1}{2(k_BT)^3}\int_0^\infty k_3(E_c)E_c^2e^{-E_c/(k_BT)}dE_c,
\end{equation}

\noindent where $k_B$ is the Boltzmann constant. The changes in the thermal averaged rates with temperature for the molecule and molecular ion species are presented in Fig. \ref{fig:thavg}. Comparing the data in Fig. \ref{fig:thavg} to those in Fig. \ref{fig:k3_vs_en}, it is first noticeable how the trends follow different power laws. This expected given the effect of  thermal averaging on the contribution of different collision energies at a given temperature $T$. Hence, the bending at the dissociation energy washes out and becomes recognizable to a smaller degree, while the power laws describing the trends shift to higher exponent values.  

The effect of the mass of the participating species on the dynamics of the recombination process is beautifully exposed by the 3-body approach we adopt and is majorly defined by Eq.~(\ref{eqn:rate}). Precisely, Fig. \ref{fig:thavg} reveals a factor of $\sqrt2$ difference between the magnitudes of the rates for hydrogen and deuterium, which agrees with the prediction of Eq.~(\ref{eqn:rate}).

In fusion divertor plasmas, in the detachment divertor regimen, where $T\sim 1$~eV, $k_3(T)\sim 10^{-32}$cm$^6$/s (see Fig.\ref{fig:thavg}), and assuming a gas density $\rho \sim 10^{15}$ cm$^{-3}$~\cite{Alacator} the collisional rate 

\begin{equation}
\Gamma=k_3(T)\rho^2,    
\end{equation}
is $\sim 0.01$~s$^{-1}$. Therefore, ion-atom-atom three-body recombination may not play a relevant role in volumentric plasma recombination processes. However, ion-atom-atom three-body recombination has a smoother temperature dependence than the most prominent three-body recombination process in volumetric recombination: e + e + H$^+\rightarrow$ H + e or e + e + D$^+\rightarrow$ D + e, which goes like $T^{-9/2}$~\cite{Sayasov}. Therefore, at higher plasma temperatures ion-atom-atom three-body recombination may start to play a relevant role in the plasma dynamics.

\section{Summary and conclusions}

This work presents a first-principles calculation of the ion-atom-atom three-body recombination rate in deuterium and hydrogen plasmas. Our approach is based on a classical trajectory calculation in hyperspherical coordinates, supported by the large number of partial waves contributing to the atom-ion scattering. The main result is the formation of molecular ions, i.e., H$_2^+$ and D$_2^+$ in cold hydrogen and deuterium plasmas, respectively, to the detriment of neutral species. Furthermore, we find a specific change of trend in the three-body recombination rate as a function of collision energy around the dissociation energy of the molecular ion, which correlates with the role of short-range physics in the reaction dynamics. 

The collisional rate associated with three-body recombination is $\sim 0.01$~s$^{-1}$ in typical fusion divertor plasmas conditions, which is a relatively slow process compared with other volumetric recombination mechanisms. However, the situation may change at higher temperatures since ion-atom-atom three-body recombination shows a weaker temperature dependence than other three-body volumetric recombination processes. Finally, it is worth mentioning that the current approach could be extended to treat more involved three-body processes like H$_2^+$ + H + H$\rightarrow$ H$_3^+$ + H, which may play a significant role in plasma physics.

\section{Acknowledgments}

The authors thank Dr. V. Kokoouline for pointing out ion-neutral collisions' role in plasmas and identifying the role of spin degrees of freedom in the H$_3^+$ potential energy surface. M. M and J. P.-R. thanks the support of the Deutsche Forschungsgemeinschaft (DFG – German Research Foundation) under the grant number PE 3477/2 - 493725479.



\bibliography{main}

\end{document}